\documentclass[12pt, a4paper]{article}
\usepackage[english]{babel}
\usepackage[utf8]{inputenc}

\usepackage{amsmath}
\usepackage{amssymb}
\usepackage{amsfonts}
\usepackage{amsbsy}

\usepackage{verbatim}
\usepackage{bm}

\usepackage{graphicx}

\usepackage{newlfont}
\usepackage{float}
\usepackage[paperwidth=192mm,
			paperheight=262mm,
			vmargin={19mm,19mm},
			hmargin={13.7mm,13.7mm},
			headsep=12pt,
			footskip=12pt]{geometry}

\usepackage{hyperref}

\DeclareMathOperator{\grad}{\nabla}
\DeclareMathOperator{\dive}{\nabla\cdot}

\usepackage{siunitx}

\usepackage{subcaption}

\begin{document}

\title{A comparison of different approaches to compute surface tension contribution in incompressible two-phase flows}

\author{Giuseppe Orlando$^{(1)}$ \\ 
		Paolo Francesco Barbante$^{(2)}$, Luca Bonaventura$^{(2)}$}

\date{}

\maketitle

\begin{center}
{
\small
$^{(1)}$ CMAP, CNRS, \'{E}cole polytechnique, Institut Polytechnique de Paris \\ Route de Saclay, 91120 Palaiseau, France \\
{\tt giuseppe.orlando@polytechnique.edu} \\
$^{(2)}$  
MOX - Dipartimento di Matematica, Politecnico di Milano \\
Piazza Leonardo da Vinci 32, 20133 Milano, Italy \\
{\tt paolo.barbante@polimi.it, luca.bonaventura@polimi.it}
}
\end{center}


\vspace*{0.5cm}

\pagebreak

\abstract{We perform a quantitative assessment of different strategies to compute the contribution due to surface tension in incompressible two-phase flows using a conservative level set (CLS) method. More specifically, we compare classical approaches, such as the direct computation of the curvature from the level set or the Laplace-Beltrami operator, with an evolution equation for the mean curvature recently proposed in literature. We consider the test case of a static bubble, for which an exact solution for the pressure jump across the interface is available, and the test case of an oscillating bubble, showing pros and cons of the different approaches.}

\pagebreak

\section{Introduction}
\label{sec:introd} \indent

Interfacial flows with surface tension play an important role in several industrial and engineering applications \cite{keller:1983, levich:1969}. Many modelling approaches have been proposed to capture the motion of the interface. We consider here the conservative level set (CLS) method, originally proposed in \cite{olsson:2005}, \cite{olsson:2007}, to which we refer for a detailed description of the scheme. In this framework, the normal to the interface and the curvature are implicitly determined from the level set function. We compare here different approaches to compute the force due to surface tension. More specifically, we consider three possible strategies: the use of the Laplace-Beltrami operator, the estimation of the total curvature directly from the level set function, and the use of an evolution equation for the mean curvature recently proposed in \cite{orlando:2023}.

The paper is structured as follows: in Section \ref{sec:model}, we briefly recall the different formulations chosen to model the surface tension force. In Section \ref{sec:num_method}, we briefly outline the numerical method employed for the analysis. Section \ref{sec:num_results} is devoted to a quantitative assessment of relations introduced in Section \ref{sec:model} for the test case of a static bubble and for the test case of an oscillating bubble. Finally, some conclusions and insights for future work are presented in Section \ref{sec:conclu}.

\section{Mathematical model}
\label{sec:model} \indent

Let $\Omega \subset \mathbb{R}^{d}, 2 \le d \le 3$ be a connected open bounded set with a sufficiently smooth boundary $\partial\Omega$ and denote by $\mathbf{x}$ the spatial coordinates and by $t$ the temporal coordinate. The two fluids in $\Omega$ are considered immiscible and they are contained in the subdomains $\Omega_{1}(t)$ and $\Omega_{2}(t)$, respectively, so that $\overline{\Omega_{1}(t)} \cup \overline{\Omega_{2}(t)} = \overline{\Omega}$. The interface between the two fluids is denoted by $\Gamma(t)$, defined as $\Gamma(t) = \partial\Omega_{1}(t) \cap \partial\Omega_2(t)$.
We consider the classical unsteady, isothermal, incompressible Navier-Stokes equations without gravity, which read as follows \cite{orlando:2024}:
\begin{eqnarray} \label{eq:ns_incomp}
	\frac{\partial\left(\rho(\mathbf{x}, t)\mathbf{u}\right)}{\partial t} + \dive\left(\rho(\mathbf{x}, t)\mathbf{u} \otimes \mathbf{u}\right) &=& -\nabla p + \dive\left[2\mu(\mathbf{x}, t)\mathbf{D}(\mathbf{u})\right] + \mathbf{f}_{\sigma} \nonumber \\
	\dive\mathbf{u} &=& 0,   
\end{eqnarray}
for $\mathbf{x} \in \Omega$, $t \in (0, T_{f}]$, supplied with suitable initial and boundary conditions. Here, $T_{f}$ is the final time, $\mathbf{u}$ is the fluid velocity, $p$ is the pressure, $\rho$ is the fluid density and $\mu$ is the dynamic viscosity. We assume that both density and viscosity are defined as
\begin{equation}
	\rho(\mathbf{x}, t) = 
	\begin{cases}
		\rho_{1} \quad \text{in } \Omega_{1}(t) \\
		\rho_{2} \quad \text{in } \Omega_{2}(t) \\
	\end{cases}
	\qquad \text{and} \qquad
	\mu(\mathbf{x},t) = 
	\begin{cases}
		\mu_{1} \quad \text{in } \Omega_{1}(t) \\
		\mu_{2} \quad \text{in } \Omega_{2}(t) \\
	\end{cases}
\end{equation}
with $\rho_{1}, \rho_{2}, \mu_{1},$ and $\mu_{2}$ constant values. Moreover, $\mathbf{D}(\mathbf{u})$ denotes the symmetric part of the gradient. Finally, $\mathbf{f}_{\sigma}$ represents a volumetric force which takes into account surface tension, defined as \cite{hysing:2009}
\begin{equation}
	\mathbf{f}_{\sigma} = \sigma\kappa\mathbf{n}_{\Gamma}\delta(\Gamma(t)),
\end{equation} 
where $\sigma$ is the constant surface tension coefficient, $\mathbf{n}_{\Gamma}$ is the outward unit normal to $\Gamma$, $\kappa = -\dive\mathbf{n}_{\Gamma}$ is the total curvature, and $\delta(\Gamma(t))$ is the Dirac delta distribution supported on the interface. In the following, for the sake of simplicity in the notation, we omit the explicit dependence on space and time for the different quantities. As discussed in \cite{lafaurie:1994}, we can rewrite the volumetric force as
\begin{equation}
	\mathbf{f}_{\sigma} = \dive\left[\sigma\left(\mathbf{I} - \mathbf{n}_{\Gamma} \otimes \mathbf{n}_{\Gamma}\right)\delta(\Gamma)\right],
\end{equation}
which corresponds to the application of the Laplace-Beltrami operator. The CLS method \cite{olsson:2005, olsson:2007} describes implicitly the interface in terms of a regularized Heaviside function $\phi$ and its evolution equation reads as follows:
\begin{equation}
	\frac{\partial\phi}{\partial t} + \mathbf{u} \cdot \nabla\phi = 0.
\end{equation}
For the sake of completeness, we report the definition of $\phi$ in terms of the signed distance function $\varphi$, employed in the classical level set method \cite{osher:2005}. The following relation is assumed:
\begin{equation}\label{eq:conservative_level_set_function}
	\phi = \frac{1}{1 + e^{-\varphi/\varepsilon}},
\end{equation}
where $\varepsilon$ helps smoothing the transition of the discontinuous physical properties between the two subdomains and it is also known as interface thickness.
From definition \eqref{eq:conservative_level_set_function}, it follows that
\begin{equation}
	\Gamma = \left\{\mathbf{x} \in \overline{\Omega} : \phi = \frac{1}{2}\right\}.
\end{equation}
The Continuum Surface Force (CSF) approach, introduced in \cite{brackbill:1992}, is employed to treat density, viscosity, and surface tension. More specifically, we set
\begin{eqnarray}
	\rho &\approx& \rho_{2} + \left(\rho_{1} - \rho_{2}\right)\phi \\ 
	\mu &\approx& \mu_{2} + \left(\mu_{1} - \mu_{2}\right)\phi. 
\end{eqnarray}
For the surface tension term, we recall that \cite{estrada:1980}
\begin{equation}
	\delta(\Gamma) = \delta\left(\phi - \frac{1}{2}\right)\left|\nabla\phi\right|,
\end{equation}
where $\delta\left(\phi - \frac{1}{2}\right)$ is the Dirac delta distribution with support equal to the interface implicitly described by $\phi = \frac{1}{2}$. Hence, we consider the following approximations:
\begin{eqnarray}
	\mathbf{f}_{\sigma} &\approx& \dive\left[\sigma\left(\left|\nabla\phi\right| - \frac{\nabla\phi \otimes \nabla\phi}{\left|\nabla\phi\right|}\right)\right] \label{eq:surface_tension_laplace_beltrami} \\
	\mathbf{f}_{\sigma} &\approx& -\sigma\dive\left(\frac{\nabla\phi}{\left|\nabla\phi\right|}\right)\nabla\phi = \sigma\frac{\nabla\phi \cdot \mathbf{H}_{\phi}\nabla\phi - \left|\nabla\phi\right|^{2}\Delta\phi}{\left|\nabla\phi\right|^{3}} \label{eq:surface_tension_level_set}
\end{eqnarray}
where we exploit the relation $\mathbf{n}_{\Gamma} = \frac{\nabla\phi}{\left|\nabla\phi\right|}$. Notice that, since $\phi$ represents a regularized Heaviside function, we approximate 
\begin{equation}
	\delta\left(\phi - \frac{1}{2}\right) = \frac{d\theta\left(\phi - \frac{1}{2}\right)}{d\phi} \approx \frac{d\theta\left(\theta\left(\varphi\right) - \frac{1}{2}\right)}{d\phi} = \frac{d\theta\left(\varphi\right)}{d\phi} \approx \frac{d\phi}{d\phi} = 1
\end{equation}
with $\theta$ denoting the Heaviside function. The second relation in \eqref{eq:surface_tension_level_set} is based on the so-called Bonnet's formula \cite{henneaux:2023} for the curvature and $\mathbf{H}_{\phi}$ denotes the Hessian matrix of $\phi$. Finally, we propose here another strategy to evaluate $\mathbf{f}_{\sigma}$. We compute $\kappa$ from an evolution equation for the mean curvature $H = \frac{\kappa}{2}$ recently proposed in \cite{orlando:2023}:
\begin{eqnarray}\label{eq:mean_curvature_evolution}
	\frac{\partial H}{\partial t} + \mathbf{u} \cdot \nabla H &=& H\left(\nabla\mathbf{u}\right)\mathbf{n}_{\Gamma} \cdot \mathbf{n}_{\Gamma} + \frac{1}{2}\nabla\mathbf{n}_{\Gamma} : \left(\nabla\mathbf{u}\right)^{T} \nonumber \\
	&-& \frac{1}{2}\left(\nabla\mathbf{u}\right)^{T}\mathbf{n}_{\Gamma} \cdot \left(\nabla\mathbf{n}_{\Gamma}\right)\mathbf{n}_{\Gamma} \nonumber \\
	&-& \frac{1}{2}\left(\mathbf{n}_{\Gamma} \otimes \mathbf{n}_{\Gamma} - \mathbf{I}\right) : \left[\nabla\left[\left(\nabla\mathbf{u}\right)^{T}\mathbf{n}_{\Gamma}\right]\right]^{T},
\end{eqnarray}
or, equivalently,
\begin{equation}\label{eq:mean_curvature_evolution_bis}
	\frac{\partial H}{\partial t} + \mathbf{u} \cdot \nabla H = \frac{1}{2}\nabla\mathbf{n}_{\Gamma} : \left(\nabla\mathbf{u}\right)^{T} + \frac{1}{2}\dive\left[\left(\mathbf{I} - \mathbf{n}_{\Gamma} \otimes \mathbf{n}_{\Gamma}\right)\left(\nabla\mathbf{u}\right)^{T}\mathbf{n}_{\Gamma}\right],
\end{equation}
so that
\begin{equation}\label{eq:surface_tension_evolution_equation}
	\mathbf{f}_{\sigma} \approx 2\sigma H\nabla\phi.
\end{equation}
\\
The discretization of incompressible Navier-Stokes equations poses several major computational issues. In particular, the velocity $\mathbf{u}$ and the pressure $p$ are coupled by the incompressibility constraint $\dive\mathbf{u} = 0$. We adopt here the so-called artificial compressibility formulation, originally introduced in \cite{chorin:1967} and employed in \cite{bassi:2022}, \cite{bassi:2006}, \cite{manzanero:2020}, \cite{orlando:2024}, \cite{orlando:2022} among many others. The incompressibility constraint is relaxed and a time evolution equation for the pressure is introduced. Hence, the final form of the system reads as follows:
\begin{eqnarray}
	\frac{\partial\left(\rho\mathbf{u}\right)}{\partial t} + \dive\left(\rho\mathbf{u} \otimes \mathbf{u}\right) &=& -\nabla p + \dive\left[2\mu\mathbf{D}(\mathbf{u})\right] + \mathbf{f}_{\sigma} \nonumber \\
	\frac{1}{\rho_{0}c^{2}}\frac{\partial p}{\partial t} + \dive\mathbf{u}
	&=& 0 \\
	\frac{\partial\phi}{\partial t} + \mathbf{u} \cdot \nabla\phi &=& 0, \nonumber  
\end{eqnarray}
with $c$ being the artificial speed of sound and $\rho_{0}$ being a reference density. A dimensional analysis can be carried out (we refer to \cite{orlando:2024} for all the details), so as to obtain the following system of equations:
\begin{eqnarray}\label{eq:ns_art_comp_surf_tens_levset_cons_adim}
	\frac{\partial\left(\rho\mathbf{u}\right)}{\partial t} + \dive \left(\rho\mathbf{u} \otimes \mathbf{u}\right) &=& -\nabla p + \frac{1}{Re} \dive\left[2\mu\mathbf{D}(\mathbf{u})\right] + \frac{1}{We}\mathbf{f}_{\sigma} \nonumber \\
	M^{2}\frac{\partial p}{\partial t} + \dive \mathbf{u} &=& 0 \\
	\frac{\partial\phi}{\partial t} + \mathbf{u} \cdot \nabla\phi &=& 0, \nonumber  
\end{eqnarray} 
where, with a slight abuse of notation, we employ the same symbols to mark non-dimensional quantities. Here $Re, We$ and $M$ are the Reynolds, Weber, and Mach number, respectively, defined as
\begin{equation}
	Re = \frac{\rho_{ref}U_{ref}L_{ref}}{\mu_{ref}} \qquad We = \frac{\rho_{ref}U_{ref}^{2}L_{ref}}{\sigma} \qquad M = \frac{U_{ref}}{c},
\end{equation}
with $U_{ref}$ being the reference velocity, $L_{ref}$ denoting the reference length, $\rho_{ref}$ being the reference density and $\mu_{ref}$ denoting the reference viscosity.

\section{Numerical method}
\label{sec:num_method} \indent

In this Section, we briefly outline the numerical method employed for the discretization of system \eqref{eq:ns_art_comp_surf_tens_levset_cons_adim}. We refer to \cite{orlando:2024} for a detailed description of the numerical scheme. We consider a decomposition of the domain $\Omega$ into a family of hexahedra (quadrilaterals in the two-dimensional case) $\mathcal{T}_{h}$ and denote each element by $K$. The skeleton $\mathcal{E}$ denotes the set of all element faces and $\mathcal{E} = \mathcal{E}^{I} \cup \mathcal{E}^{B}$, where $\mathcal{E}^{I}$ is the subset of interior faces and $\mathcal{E}^{B}$ is the subset of boundary faces. Classical jump and average operators are then defined as customary for finite element discretizations. A face $e \in \mathcal{E}^{I}$ shares two elements that we denote by $K^{+}$ with outward unit normal $\mathbf{n}^{+}$ and $K^{-}$ with outward unit normal $\mathbf{n}^{-}$, whereas for a face $e \in \mathcal{E}^{B}$ we denote by $\mathbf{n}$ the outward unit normal. For a scalar function $\Psi$ the jump is defined as
\begin{equation}
	\left[\left[\Psi\right]\right] = \Psi^{+}\mathbf{n}^{+} + \Psi^{-}\mathbf{n}^{-} \quad \text{if } e \in \mathcal{E}^{I} \qquad \left[\left[\Psi\right]\right] = \Psi\mathbf{n} \quad \text{if } e \in \mathcal{E}^{B}.
\end{equation}
The average is defined as
\begin{equation}
	\left\{\left\{\Psi\right\}\right\} = \frac{1}{2}\left(\Psi^{+} + \Psi^{-}\right) \quad \text{if } e \in \mathcal{E}^{I} \qquad \left\{\left\{\Psi\right\}\right\} = \Psi \quad \text{if } e \in \mathcal{E}^{B}.	
\end{equation}
Similar definitions apply for a vector function $\boldsymbol{\Psi}$:
\begin{align}
	&\left[\left[\boldsymbol{\Psi}\right]\right] = \boldsymbol{\Psi}^{+} \cdot \mathbf{n}^{+} + \boldsymbol{\Psi}^{-} \cdot \mathbf{n}^{-} \quad \text{if } e \in \mathcal{E}^{I} \qquad 
	\left[\left[\boldsymbol{\Psi}\right]\right] = \boldsymbol{\Psi} \cdot \mathbf{n} \quad \text{if } e \in \mathcal{E}^{B} \\
	&\left\{\left\{\boldsymbol{\Psi}\right\}\right\} = \frac{1}{2}\left(\boldsymbol{\Psi}^{+} + \boldsymbol{\Psi}^{-}\right) \quad \text{if } e \in \mathcal{E}^{I} \qquad \left\{\left\{\boldsymbol{\Psi}\right\}\right\} = \boldsymbol{\Psi} \quad \text{if } e \in \mathcal{E}^{B}.
\end{align}
We now introduce the following finite element spaces:
\begin{equation}
	Q_{k} = \left\{v \in L^{2}(\Omega) : v\rvert_{K} \in \mathbb{Q}_{k} \quad \forall K \in \mathcal{T}_{h}\right\}
\end{equation} 
and
\begin{equation}
	\mathbf{Q}_{k} = \left[Q_{k}\right]^d,
\end{equation}
where $\mathbb{Q}_{k}$ is the space of polynomials of degree $k$ in each coordinate direction. The finite element spaces that will be used for the discretization of velocity and pressure are $\mathbf{V}_{h} = \mathbf{Q}_{k}$ and $W_{h} = Q_{k-1} \cap L^2_{0}(\Omega)$, respectively, where $k \ge 2$. For what concerns the level set function and the curvature, we consider instead $X_{h} = Q_{r}$ with $r \ge 2$, so that its gradient is at least a piecewise linear polynomial. In the present work, we consider $k = r = 2$. 

We briefly recall for the convenience of the reader the formulation of the TR-BDF2. Let $\Delta t = T_{f}/N$ be a discrete time step and $t^{n} = n\Delta t, n = 0, \dots, N$, be discrete time levels for a generic time dependent problem $\bm{u}^{\prime} = \mathcal{N}(\bm{u})$. The incremental form of the TR-BDF2 scheme can be described in terms of two stages, the first one from $t^{n}$ to $t^{n+\gamma} = t^{n} + \gamma\Delta t$, and the second one from $t^{n+\gamma}$ to $t^{n+1}$, as follows:
\begin{eqnarray}
	\frac{\bm{u}^{n+\gamma} - \bm{u}^{n}}{\gamma\Delta t} &=& \frac{1}{2}\mathcal{N}\left(\bm{u}^{n+\gamma}\right) + \frac{1}{2}\mathcal{N}\left(\bm{u}^{n}\right) \\
	\frac{\bm{u}^{n+1} - \bm{u}^{n+\gamma}}{\left(1 - \gamma\right)\Delta t} &=& \frac{1}{2 - \gamma}\mathcal{N}\left(\bm{u}^{n+1}\right) + \frac{1 - \gamma}{2\left(2 - \gamma\right)}\mathcal{N}\left(\bm{u}^{n+\gamma}\right) + \frac{1 - \gamma}{2\left(2 - \gamma\right)}\mathcal{N}\left(\bm{u}^{n}\right).
\end{eqnarray}
Here, $\bm{u}^{n}$ denotes the approximation at time $n = 0, \dots, N$. Notice that, in order to guarantee L-stability, one has to choose $\gamma = 2 - \sqrt{2}$ \cite{hosea:1996}. 

A reinitialization procedure is adopted for the level set function. We employ the following PDE \cite{olsson:2005, olsson:2007}:
\begin{equation}
	\frac{\partial\phi}{\partial\tau} + \dive\left(u_{c}\phi\left(1 - \phi\right)\mathbf{n}_{\Gamma}\right) = \dive\left(\beta\varepsilon u_{c}\left(\nabla\phi \cdot \mathbf{n}_{\Gamma}\right)\mathbf{n}_{\Gamma}\right), 
\end{equation}
where $\tau$ is an artificial pseudo-time variable, $u_{c}$ is an artificial compression velocity, and $\beta$ is a constant. It is important to notice that $\mathbf{n}_{\Gamma}$ does not change during the reinitialization procedure, but it is computed using the initial value of the level set function.
The supplementary equation for the mean curvature \eqref{eq:mean_curvature_evolution}-\eqref{eq:mean_curvature_evolution_bis} has to be discretized using \eqref{eq:surface_tension_evolution_equation} to evaluate the surface tension contribution. For the sake of completeness, we report the first stage of the TR-BDF2 scheme for \eqref{eq:mean_curvature_evolution}-\eqref{eq:mean_curvature_evolution_bis}:
\begin{eqnarray}\label{eq:mean_curvature_evolution_bis_first_stage}
	&&\frac{H^{n+\gamma} - H^{n}}{\gamma \Delta t} + \frac{1}{2}\mathbf{u}^{n + \frac{\gamma}{2}} \cdot \nabla H^{n+\gamma} + \frac{1}{2}\mathbf{u}^{n + \frac{\gamma}{2}} \cdot \nabla H^{n} = \nonumber \\ &&\frac{1}{2}\left(\frac{1}{2}\nabla\mathbf{n}_{\Gamma}^{n+\gamma} : \nabla\mathbf{u}^{n + \frac{\gamma}{2}}\right) + \frac{1}{2}\left(\frac{1}{2}\nabla\mathbf{n}_{\Gamma}^{n} : \nabla\mathbf{u}^{n}\right) \\
	&+&\frac{1}{2}\left(\frac{1}{2}\dive\left[\left(\mathbf{I} - \mathbf{n}_{\Gamma}^{n+\gamma} \otimes \mathbf{n}_{\Gamma}^{n+\gamma}\right)\left(\nabla\mathbf{u}^{n + \frac{\gamma}{2}}\right)^{T}\mathbf{n}_{\Gamma}^{n+\gamma}\right]\right) \nonumber \\
	&+&\frac{1}{2}\left(\frac{1}{2}\dive\left[\left(\mathbf{I} - \mathbf{n}_{\Gamma}^{n} \otimes \mathbf{n}_{\Gamma}^{n}\right)\left(\nabla\mathbf{u}^{n}\right)^{T}\mathbf{n}_{\Gamma}^{n}\right]\right), \nonumber
\end{eqnarray}
where $\mathbf{u}^{n + \frac{\gamma}{2}} = \left(1 + \frac{\gamma}{2\left(1-\gamma\right)}\right)\mathbf{u}^{n} - \frac{\gamma}{2\left(1-\gamma\right)}\mathbf{u}^{n-1}$ is defined by extrapolation. Hence, the weak formulation for \eqref{eq:mean_curvature_evolution_bis_first_stage} reads as follows:
\begin{eqnarray}
	&&\sum_{K \in \mathcal{T}_{h}}\int_{K} \frac{H^{n+\gamma}}{\gamma\Delta t}w d\Omega + \frac{1}{2}\sum_{K \in \mathcal{T}_{h}}\int_{K} \mathbf{u}^{n + \frac{\gamma}{2}} \cdot \nabla H^{n+\gamma} w d\Omega \nonumber \\
	&+& \frac{1}{2}\sum_{e \in \mathcal{E}} \int_{e} \left\{\left\{H^{n+\gamma}\mathbf{u}^{n + \frac{\gamma}{2}}\right\}\right\} \cdot \left[\left[w\right]\right]d\Sigma - \frac{1}{2}\sum_{e \in \mathcal{E}} \int_{e} \left\{\left\{\mathbf{u}^{n + \frac{\gamma}{2}}\right\}\right\} \cdot \left[\left[H^{n+\gamma}w\right]\right]d\Sigma \nonumber \\
	&+& \frac{1}{2}\sum_{e \in \mathcal{E}} \int_{e} \frac{\lambda^{n + \frac{\gamma}{2}}}{2}\left[\left[H^{n+\gamma}\right]\right] \cdot \left[\left[w\right]\right]d\Sigma \\
	&=& \sum_{K \in \mathcal{T}_{h}}\int_{K} \frac{H^{n}}{\gamma \Delta t}w d\Omega + \frac{1}{2}\sum_{K \in \mathcal{T}_{h}}\int_{K} \mathbf{u}^{n + \frac{\gamma}{2}} \cdot \nabla H^{n} w d\Omega \nonumber \\
	&-& \frac{1}{2}\sum_{e \in \mathcal{E}} \int_{e} \left\{\left\{H^{n}\mathbf{u}^{n + \frac{\gamma}{2}}\right\}\right\} \cdot \left[\left[w\right]\right]d\Sigma - \frac{1}{2}\sum_{e \in \mathcal{E}} \int_{e} \left\{\left\{\mathbf{u}^{n + \frac{\gamma}{2}}\right\}\right\} \cdot \left[\left[H^{n}w\right]\right]d\Sigma \nonumber \\
	&-& \frac{1}{2}\sum_{e \in \mathcal{E}} \int_{e} \frac{\lambda^{n + \frac{\gamma}{2}}}{2}\left[\left[H^{n}\right]\right] \cdot \left[\left[w\right]\right]d\Sigma \nonumber \\
	&+& \frac{1}{2}\sum_{K \in \mathcal{T}_{h}}\int_{K}\frac{1}{2}\nabla\mathbf{n}_{\Gamma}^{n+\gamma} : \nabla\mathbf{u}^{n + \frac{\gamma}{2}}wd\Omega + \frac{1}{2}\sum_{K \in \mathcal{T}_{h}}\int_{K}\frac{1}{2}\nabla\mathbf{n}_{\Gamma}^{n} : \nabla\mathbf{u}^{n}wd\Omega \nonumber \\
	&+& \frac{1}{2}\sum_{e \in \mathcal{E}} \int_{e} \left\{\left\{\nabla\mathbf{n}_{\Gamma}^{n+\gamma}\mathbf{u}^{n + \frac{\gamma}{2}}\right\}\right\} \cdot \left[\left[w\right]\right]d\Sigma + \frac{1}{2}\sum_{e \in \mathcal{E}} \int_{e} \left\{\left\{\nabla\mathbf{n}_{\Gamma}^{n}\mathbf{u}^{n}\right\}\right\} \cdot \left[\left[w\right]\right]d\Sigma \nonumber \\
	&-& \frac{1}{2}\sum_{e \in \mathcal{E}} \int_{e} \left\{\left\{\nabla\mathbf{n}_{\Gamma}^{n+\gamma}\mathbf{u}^{n + \frac{\gamma}{2}}\right\}\right\} : \left<\left<\mathbf{u}^{n + \frac{\gamma}{2}}w\right>\right>d\Sigma - \frac{1}{2}\sum_{e \in \mathcal{E}} \int_{e} \left\{\left\{\nabla\mathbf{n}_{\Gamma}^{n}\right\}\right\} : \left<\left<w\mathbf{u}^{n}\right>\right>d\Sigma \nonumber \\
	&-& \frac{1}{2}\sum_{K \in \mathcal{T}_{h}} \int_{K}\frac{1}{2} \left(\mathbf{I} - \mathbf{n}_{\Gamma}^{n+\gamma} \otimes \mathbf{n}_{\Gamma}^{n+\gamma}\right)\left(\nabla\mathbf{u}^{n + \frac{\gamma}{2}}\right)^{T}\mathbf{n}_{\Gamma}^{n+\gamma} \cdot \nabla w d\Omega \nonumber \\
	&+& \frac{1}{2}\sum_{e \in \mathcal{E}} \int_{e} \left\{\left\{\left(\mathbf{I} - \mathbf{n}_{\Gamma}^{n+\gamma} \otimes \mathbf{n}_{\Gamma}^{n+\gamma}\right)\left(\nabla\mathbf{u}^{n + \frac{\gamma}{2}}\right)^{T}\mathbf{n}_{\Gamma}^{n+\gamma}\right\}\right\} \cdot \left[\left[w\right]\right] \nonumber \\
	&-& \frac{1}{2}\sum_{K \in \mathcal{T}_{h}} \int_{K}\frac{1}{2} \left(\mathbf{I} - \mathbf{n}_{\Gamma}^{n} \otimes \mathbf{n}_{\Gamma}^{n}\right)\left(\nabla\mathbf{u}^{n}\right)^{T}\mathbf{n}_{\Gamma}^{n} \cdot \nabla w d\Omega \nonumber \\
	&+& \frac{1}{2}\sum_{e \in \mathcal{E}} \int_{e} \left\{\left\{\left(\mathbf{I} - \mathbf{n}_{\Gamma}^{n} \otimes \mathbf{n}_{\Gamma}^{n}\right)\left(\nabla\mathbf{u}^{n}\right)^{T}\mathbf{n}_{\Gamma}^{n}\right\}\right\} \cdot \left[\left[w\right]\right] 
	\qquad \forall w \in X_{h}, \nonumber
\end{eqnarray}
with
\begin{equation}
	\lambda^{n + \frac{\gamma}{2}} = \max\left(\left|\left(\mathbf{u}^{n + \frac{\gamma}{2}}\right)^{+} \cdot \mathbf{n}_{e}^{+}\right|, \left|\left(\mathbf{u}^{n + \frac{\gamma}{2}}\right) \cdot \mathbf{n}_{e}^{-}\right|\right).
\end{equation}
The numerical approximation of non-conservative terms is based on the approach proposed in \cite{bassi:1997}. We recast the non-conservative term into the sum of two contributions: the first one takes into account the elementwise gradient, whereas the second one considers its jumps across the element faces. The second TR-BDF2 stage can be described in a similar manner. Finally, we consider the following spatial discretization for $\mathbf{f}_{\sigma}$:
\begin{equation}
	\mathbf{f}_{\sigma} \approx 2H\nabla\phi \approx \sum_{K \in \mathcal{T}_{h}} \int_{K} 2H\nabla\phi \cdot \boldsymbol{\varphi}d\Omega + \sum_{e \in \mathcal{E}} \int_{e} 2\left\{\left\{H\phi\right\}\right\}\left[\left[\boldsymbol{\varphi}\right]\right]d\Sigma - \sum_{e \in \mathcal{E}} \int_{e} 2\left\{\left\{H\right\}\right\}\left[\left[\phi\boldsymbol{\varphi}\right]\right]d\Sigma,
\end{equation}
which is again based on the approach presented in \cite{bassi:1997}. Here, $\boldsymbol{\varphi}$ is a basis function of the finite element space chosen to discretize the momentum equation, i.e. $\mathbf{V}_{h}$. 

\section{Numerical results}
\label{sec:num_results}

In this Section, we present the results from simulations comparing the different approaches presented in Section \ref{sec:model} to evaluate $\mathbf{f}_{\sigma}$. The numerical method outlined in Section \ref{sec:num_method} has been implemented in the framework of the $\mathit{deal.II}$ library \cite{arndt:2023, bangerth:2007}. 

\subsection{Static bubble}
\label{ssec:static_bubble}

We consider the 2D stationary bubble in a zero force field described e.g. in \cite{deshpande:2012, hysing:2006, popinet:1999}. According to the Laplace–Young law \cite{francois:2006}, the pressure jump across the interface of two immiscible fluids is directly related to surface tension. Indeed, the following relation holds:
\begin{equation}
	\Delta p = p_{in} - p_{out} = \frac{\sigma}{R},
\end{equation} 
where $p_{in}$ and $p_{out}$ are the pressure inside and outside the bubble, respectively, whereas $R$ is the radius of the bubble. A bubble with $R = \SI{0.25}{\meter}$ centered in $\left(x_{0}, y_{0}\right) = \left(0.5, 0.5\right)$ in $\Omega = \left(0,1\right)^{2}$ is considered. The fluid properties are $\rho_{1} = \rho_{2} = \SI[parse-numbers=false]{10^{4}}{\kilogram\per\meter\cubed}$, $\mu_{1} = \mu_{2} = \SI{1}{\kilogram\per\meter\per\second}$, and $\sigma = \SI{1}{\newton\per\meter}$, so that $\Delta p = \SI{4}{\newton\per\meter\squared}$. Finally, we set $T_{f} = \SI{25}{\second}$ and a fixed time step $\Delta t = \SI{0.2}{\second}$. The artificial speed of sound is set to $c \approx \SI{1428}{\meter\per\second}$, which is of the same order of magnitude of the speed of sound in water. The reference length is equal to $L_{ref} = 2R = \SI{0.5}{\meter}$ and the reference velocity $U_{ref}$ is chosen in such a way that the reference pressure $p_{ref} = \rho_{ref}U_{ref}^{2}$ is unitary. Hence, we get $U_{ref} = \SI{0.01}{\meter\per\second}$, so as to obtain $Re = 50, We = 0.5$, and $M = 7 \times 10^{-6}$. Periodic boundary conditions are considered. Following \cite{deshpande:2012}, we consider three metrics to assess the pressure jump computation:
\begin{enumerate}
	\item $\Delta p_{total} = p_{in} - p_{out}$,  where the subscripts $in$ denotes quantities inside the bubble (averaged over cells with $r = \sqrt{\left(x - x_{0}\right)^2 + \left(y- y_{0}\right)^2} \le R$)
	and $out$ quantities outside the drop (averaged over cells with $r > R$,
	\item $\Delta p_{partial} = p_{in} - p_{out}$,  where the subscripts $in$ denotes quantities inside the bubble (averaged over cells with $r \le \frac{R}{2}$
	and $out$ quantities outside the drop (averaged over cells with $r > \frac{3}{2}{R}$), so as to avoid the interface region,
	\item $\Delta p_{max} = p_{max} - p_{min}$,  where $p_{max}$ and $p_{min}$ are the maximum and minimum pressure on the whole domain, respectively.
\end{enumerate}
We also monitor the degree of circularity, defined in \cite{hysing:2009} as
\begin{equation}
	\chi = \frac{2\sqrt{\pi\left|\Omega_{2}\right|}}{P_{b}},
\end{equation}
with $\Omega_{2}$ being the subdomain occupied by the bubble, $\left|\Omega_{2}\right|$ being the area of the bubble and $P_{b}$ being its perimeter. The degree of circularity is the ratio between the perimeter
of a circle with the same area of the bubble and the current perimeter of the bubble itself. Since, for a perfectly circular bubble, the degree of circularity is unitary, we expect values equal to one or very close to it.

\subsubsection{Use of Laplace-Beltrami operator}
\label{ssec:laplace_beltrami}

\begin{figure}[h!]
	\centering
	\includegraphics[width = 0.8\textwidth]{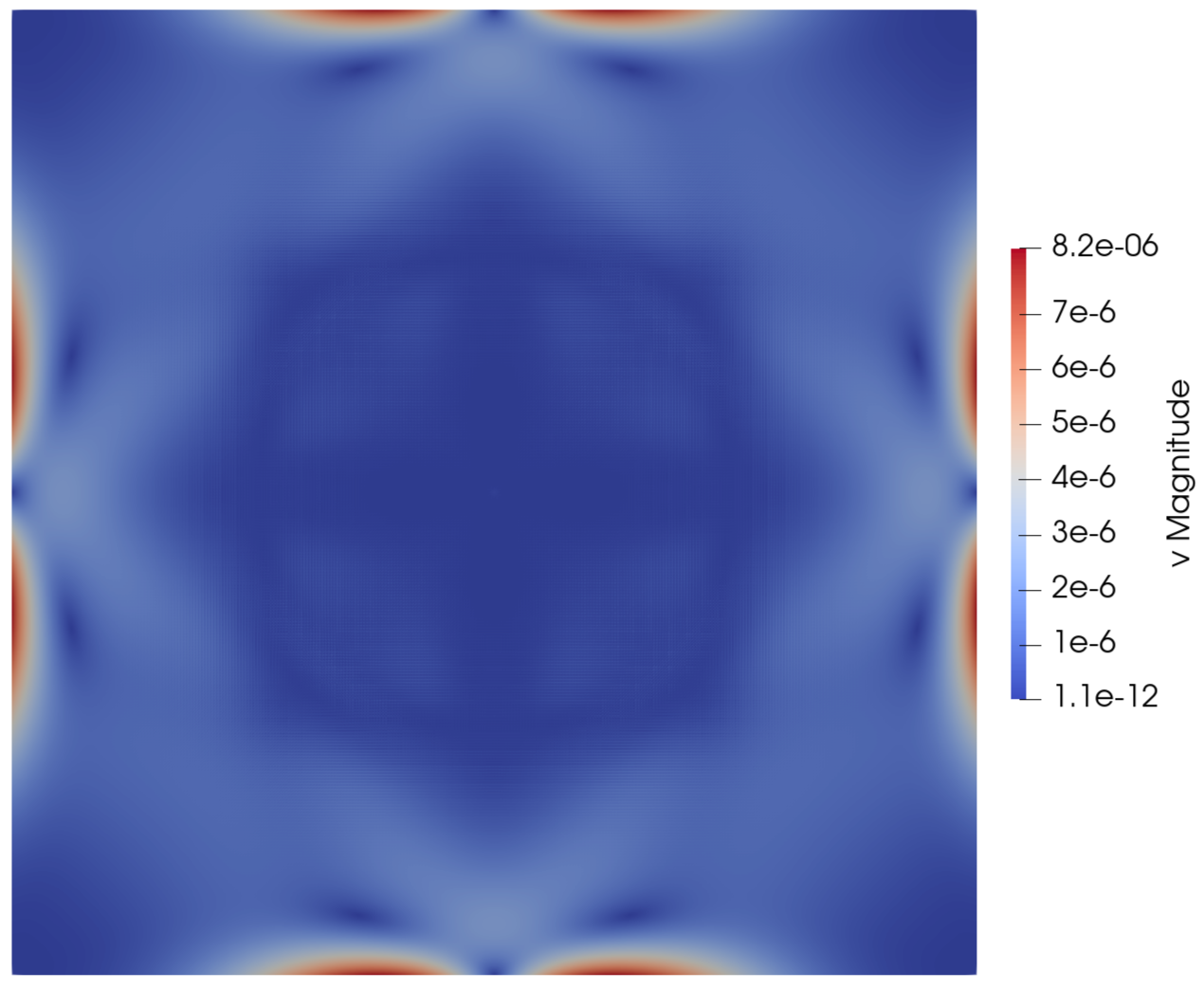}
	\caption{Static bubble test case, velocity magnitude at $t = T_{f}$ using \eqref{eq:surface_tension_laplace_beltrami} with $N_{el} = 320$.}
	\label{fig:static_bubble_spurious_currents_Laplace_Beltrami}
\end{figure}

In this Section, we consider relation \eqref{eq:surface_tension_laplace_beltrami} to evaluate $\mathbf{f}_{\sigma}$ and we report the results in Table \ref{tab:static_bubble_laplace_beltrami_results}. Notice that we add a small number $\eta = 10^{-10}$ to the denominator $\left|\nabla\phi\right|$ so as to avoid division by zero. The same workaround is adopted to evaluate the unit normal vector in \eqref{eq:surface_tension_level_set}. The approximation is very robust and no oscillations for the pressure jump arise during the time evolution (see Figure \ref{fig:static_bubble_delta_pres_partial}). However, one can easily notice that we achieve convergence towards a value which is different with respect to the analytical one. This is in agreement with what happens using the Volume of Fluid method, as reported in \cite{deshpande:2012}, and we will further discuss this issue in Section \ref{ssec:evolution_curvature}. Finally, the degree of circularity is always around $1$ for all the configurations, meaning that the circular shape is preserved. The relative error for $\chi$ with $N_{el}$ is around $2 \times 10^{-5}$. Finally, the generation of spurious currents, which typically appear in the form of vortices around the interface, is strongly reduced, as evident from Figure \ref{fig:static_bubble_spurious_currents_Laplace_Beltrami}. This further confirms the robustness of this approach.

\begin{table}[h!]
	\centering
	\begin{tabular}{|c|c|c|c|c|} 
		\hline
		$N_{el}$ & $\Delta p_{total}$ & $\Delta p_{partial}$ & $\Delta p_{max}$ & $\chi$ \\
		\hline
		$40$ & $3.37$ & $4.15$ & $4.16$ & $1.00$ \\ 
		\hline
		$80$ & $3.37$ & $4.13$ & $4.16$ & $1.00$ \\
		\hline
		$160$ & $3.37$ & $4.13$ & $4.16$ & $1.00$\\ 
		\hline
		$320$ & $3.37$ & $4.13$ & $4.17$ & $1.00$ \\ 
		\hline
	\end{tabular}
	\caption{Static bubble test case, pressure jump across the interface and degree of circularity at final time $t = T_{f}$ with $\mathbf{f}_{\sigma}$ computed using the Laplace-Beltrami operator \eqref{eq:surface_tension_laplace_beltrami}. $N_{el}$ denotes the number of elements along each direction.}
	\label{tab:static_bubble_laplace_beltrami_results}
\end{table}

\subsubsection{Computation of the total curvature from the level set}
\label{ssec:level_set}

In this Section, we consider relations \eqref{eq:surface_tension_level_set} for $\mathbf{f}_{\sigma}$. Table \ref{tab:static_bubble_levset_results_Bonnet} shows the results obtained with Bonnet's formula. One can easily notice that the results are much less accurate with respect to those obtained with the Laplace-Beltrami operator. Moreover, the values of pressure jump are strongly oscillating, as one can notice from Figure \ref{fig:static_bubble_delta_pres_partial}. The two relations in \eqref{eq:surface_tension_level_set}, namely the computation of $\mathbf{f}_{\sigma}$ with Bonnet's formula and with the divergence of the unit normal are equivalent for continuous functions. However, they yield different values once projected onto discrete functional spaces. Notice also that, to apply the first relation in \eqref{eq:surface_tension_level_set}, we need to project the unit normal vector $\frac{\nabla\phi}{\left|\nabla\phi\right|}$ onto a suitable space, so as to compute the divergence. In spite of this discrepancy, analogous results are achieved as evident from Table \ref{tab:static_bubble_levset_results_divergence} and Figure \ref{fig:static_bubble_delta_pres_partial}. The presence of spurious oscillations in the curvature field computed from the level set formulation is well known in the literature and different filtering approaches have been proposed to mitigate this issue, see e.g. \cite{henneaux:2023} for a comparison different filtering strategies.
For the degree of circularity, we notice a small degradation in the description of the interface, especially at coarse resolutions. However, good results are obtained overall. The relative error for $\chi$ with $N_{el} = 320$ is around $4 \times 10^{-4}$ for both relations \eqref{eq:surface_tension_level_set}. 

\begin{table}[h!]
	\centering
	\begin{tabular}{|c|c|c|c|c|} 
		\hline
		$N_{el}$ & $\Delta p_{total}$ & $\Delta p_{partial}$ & $\Delta p_{max}$ & $\chi$ \\
		\hline
		$40$ & $2.81$ & $3.40$ & $10.26$ & $0.98$ \\ 
		\hline
		$80$ & $3.00$ & $3.59$ & $3.64$ & $1.00$ \\
		\hline
		$160$ & $2.95$ & $3.54$ & $3.59$ & $1.00$ \\ 
		\hline
		$320$ & $2.94$ & $3.53$ & $3.58$ & $1.00$ \\
		\hline
	\end{tabular}
	\caption{Static bubble test case, pressure jump across the interface and degree of circularity at final time $t = T_{f}$ with $\mathbf{f}_{\sigma}$ computed using the Bonnet's formula \eqref{eq:surface_tension_level_set}. $N_{el}$ denotes the number of elements along each direction.}
	\label{tab:static_bubble_levset_results_Bonnet}
\end{table}

\begin{table}[h!]
	\centering
	\begin{tabular}{|c|c|c|c|c|} 
		\hline
		$N_{el}$ & $\Delta p_{total}$ & $\Delta p_{partial}$ & $\Delta p_{max}$ & $\chi$\\
		\hline
		$40$ & $2.26$ & $2.81$ & $9.62$ & $0.98$ \\ 
		\hline
		$80$ & $2.98$ & $3.57$ & $3.62$ & $1.00$ \\
		\hline
		$160$ & $2.95$ & $3.54$ & $3.59$ & $1.00$ \\ 
		\hline
		$320$ & $2.94$ & $3.53$ & $3.58$ & $1.00$ \\ 
		\hline
	\end{tabular}
	\caption{Static bubble test case, pressure jump across the interface and degree of circularity at final time $t = T_{f}$ with $\mathbf{f}_{\sigma}$ computed using the divergence of unit normal vector \eqref{eq:surface_tension_level_set}. $N_{el}$ denotes the number of elements along each direction.}
	\label{tab:static_bubble_levset_results_divergence}
\end{table}

\subsubsection{Use of an evolution equation for the curvature}
\label{ssec:evolution_curvature}

\begin{figure}[h!]
	\centering
	\includegraphics[width = 0.8\textwidth]{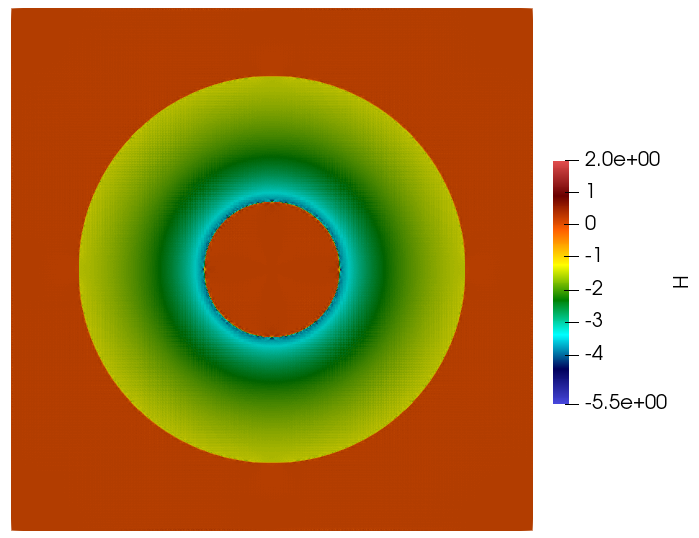}
	\caption{Static bubble test case, initialization of $H$ using \eqref{eq:mean_curvature_init} with $N_{el} = 320$.}
	\label{fig:static_bubble_curvature_init}
\end{figure}

The use of the Laplace-Beltrami operator is a very robust technique, but, as already noticed, it yields slightly inaccurate results. The reason is that there are functions, such as the unit normal vector or the curvature, whose definitions are properly meaningful only for the points on the surface $\Gamma$ \cite{nadim:1996, orlando:2023}. The capillarity flux tensor $\left(\mathbf{I} - \mathbf{n}_{\Gamma} \otimes \mathbf{n}_{\Gamma}\right)\left|\nabla\phi\right|$ introduces instead spurious contributions far from the interface. Analogous considerations hold for the total curvature computed from the level set, for which spurious or even singular values arise if $\left|\nabla\phi\right| \approx 0$. In order to overcome this issue, we employ an evolution equation for the mean curvature, so as to evaluate $\mathbf{f}_{\sigma}$ with \eqref{eq:surface_tension_evolution_equation}. This strategy is conceptually similar to the one proposed in \cite{vogl:2016}, where a curvature-augmented approach to the level set method has been proposed. However, the evolution equation \eqref{eq:mean_curvature_evolution}-\eqref{eq:mean_curvature_evolution_bis} is more general and valid for any sufficiently regular moving closed surface. A key point is the choice of the initial value for the mean curvature $H^{0}$, so as to avoid spurious contributions far from the interface. We set therefore
\begin{equation}\label{eq:mean_curvature_init}
	H^{0} = 
	\begin{cases}
		-\frac{1}{2}\frac{1}{\sqrt{\left(x - x_{0}\right)^{2} + \left(y - y_{0}\right)^{2}}} \qquad &\text{if } \left|\grad\phi^{0}\right| > \frac{\beta}{h} \\
		0 \qquad &\text{otherwise.}
	\end{cases}
\end{equation}
Here, $\beta = 5 \times 10^{-4}$, $\phi^{0}$ is the initial value of the level set function, and $h = \frac{L}{N_{el}r}$ is the space step size, with $L = \SI{1}{\meter}$ being the domain length, $N_{el}$ denoting the number of elements along each direction and $r$ being the polynomial degree of the finite element space chosen for the discretization of the mean curvature. Recall that, in the present work, we take $r = 2$. The resulting initial datum for the curvature with $N_{el} = 320$ is reported in Figure \ref{fig:static_bubble_curvature_init}. 

\begin{figure}[h!]
	\centering
	\includegraphics[width = 0.9\textwidth]{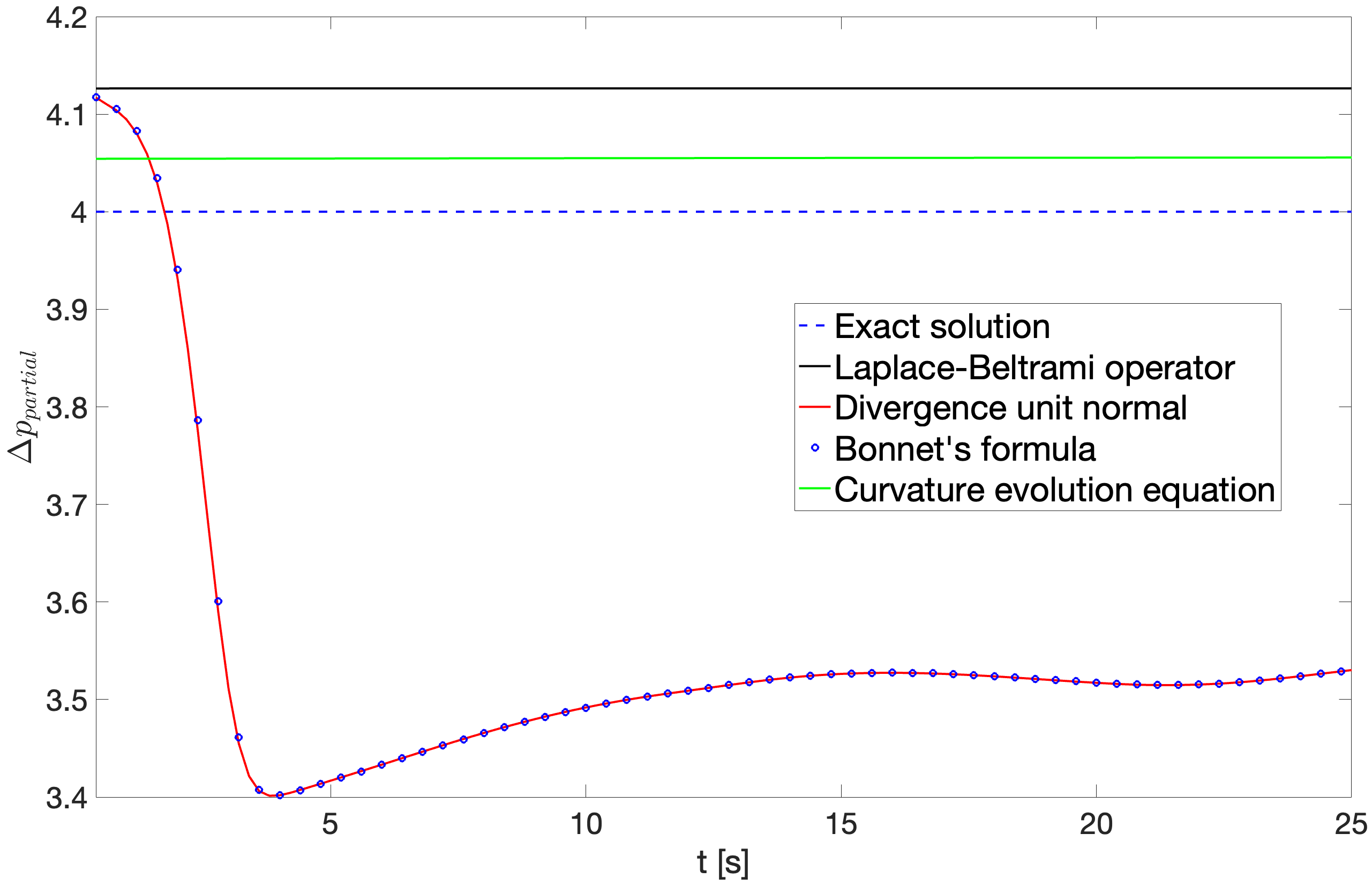}
	\caption{Static bubble test case, evolution of $\Delta p_{partial}$ with $N_{el} = 320$. The black line reports the results obtained with the Laplace-Beltrami operator \eqref{eq:surface_tension_laplace_beltrami}, the red line shows the results obtained with the Bonnet's formula in \eqref{eq:surface_tension_level_set}, whereas the blue dots represent the results achieved with the divergence of the unit normal in \eqref{eq:surface_tension_level_set}. Finally, the green line shows the results obtained with the evolution equation \eqref{eq:mean_curvature_evolution}-\eqref{eq:mean_curvature_evolution_bis} for the mean curvature and surface tension computed with \eqref{eq:surface_tension_evolution_equation}.}
	\label{fig:static_bubble_delta_pres_partial}
\end{figure}

Relation \eqref{eq:mean_curvature_init} allows us to consider contributions of the curvature in $\mathbf{f}_{\sigma}$ only at the interface and close to it, namely when $\left|\grad\phi\right|$ is above a certain threshold. Table \ref{tab:static_bubble_evolution_equation_results} shows the obtained results and one can easily notice that we achieve more accurate results with respect to those obtained with the Laplace-Beltrami operator. Moreover, $\Delta p_{partial}$ and $\Delta p_{max}$ are closer as the resolution is increased, meaning that the presence of spurious contributions far from the interface is progressively reduced. No oscillations arise for the pressure jump, as evident from Figure \ref{fig:static_bubble_delta_pres_partial}, analogously to what is observed for the Laplace-Beltrami operator. Finally, for what concerns the degree of circularity, the relative error with $N_{el} = 320$ is around $1 \times 10^{-5}$, which is analogous to the value obtained with the Laplace-Beltrami operator and one order of magnitude lower than what obtained with \eqref{eq:surface_tension_level_set}.

\begin{table}[h!]
	\centering
	\begin{tabular}{|c|c|c|c|c|} 
		\hline
		$N_{el}$ & $\Delta p_{total}$ & $\Delta p_{partial}$ & $\Delta p_{max}$ & $\chi$ \\
		\hline
		$40$ & $3.53$ & $4.36$ & $4.40$ & $0.99$ \\ 
		\hline
		$80$ & $3.38$ & $4.13$ & $4.15$ & $1.00$ \\
		\hline
		$160$ & $3.36$ & $4.10$ & $4.11$ & $1.00$ \\ 
		\hline
		$320$ & $3.35$ & $4.05$ & $4.06$ & $1.00$ \\ 
		\hline
	\end{tabular}
	\caption{Static bubble test case, pressure jump across the interface and degree of circularity at final time $t = T_{f}$ with $\mathbf{f}_{\sigma}$ computed using \eqref{eq:surface_tension_evolution_equation}. $N_{el}$ denotes the number of elements along each direction.}
	\label{tab:static_bubble_evolution_equation_results}
\end{table}

\begin{figure}[h!]
	\centering
	\begin{subfigure}{0.475\textwidth}
		\centering
		\includegraphics[width=0.9\textwidth]{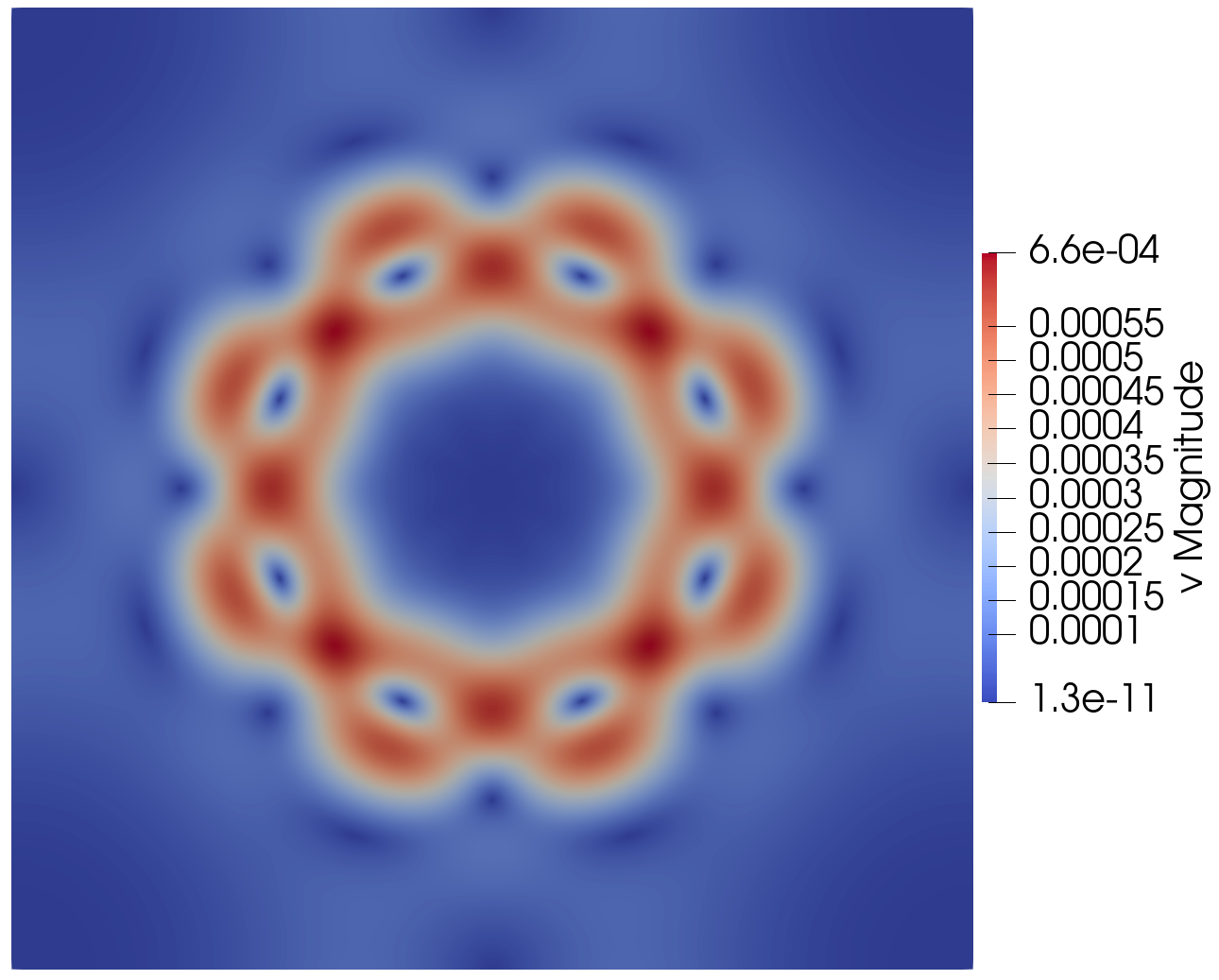} a)
	\end{subfigure}
	\begin{subfigure}{0.475\textwidth}
		\centering
		\includegraphics[width=0.9\textwidth]{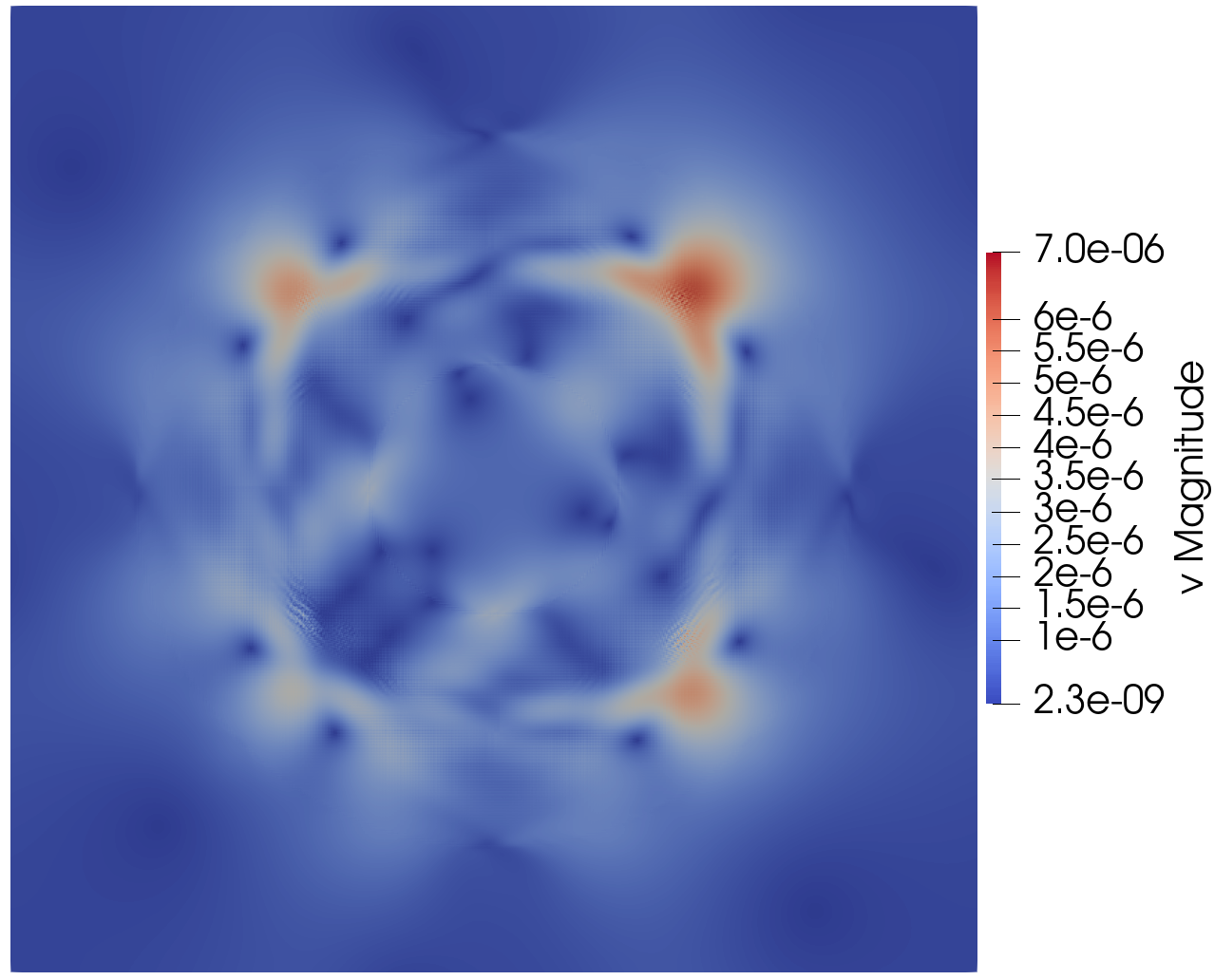} b)
	\end{subfigure}
	\caption{Static bubble test case, velocity magnitude at $t = T_{f}$ with $N_{el} = 320$, a) curvature computed from divergence of unit normal as in \eqref{eq:surface_tension_level_set}, b) curvature computed from evolution equation \eqref{eq:mean_curvature_evolution}-\eqref{eq:mean_curvature_evolution_bis} and surface tension evaluated with \eqref{eq:surface_tension_evolution_equation}.}
	\label{fig:static_bubble_spurious_currents}
\end{figure}

For longer times, the numerical solution of $H$ is corrupted. As discussed in \cite{vogl:2016}, a reinitialization procedure is necessary, analogously to what is done for the level set and, more generally, for transport equations of interfacial quantities. This issue starts arising with the formation of spurious currents in the form of vortices around the interface, as one can notice from Figure \ref{fig:static_bubble_spurious_currents}. Nevertheless, the magnitude of spurious currents is significantly reduced with respect to \eqref{eq:surface_tension_level_set}, meaning that, in this test case, the present approach is more robust and accurate with respect to the one presented in Section \ref{ssec:level_set}. The development of suitable reinitialization techniques for \eqref{eq:mean_curvature_evolution}-\eqref{eq:mean_curvature_evolution_bis} will be matter of future analysis. 

\subsection{Oscillating bubble}
\label{ssec:oscillating_bubble}

\begin{figure}[h!]
	\centering
	\includegraphics[width = 0.8\textwidth]{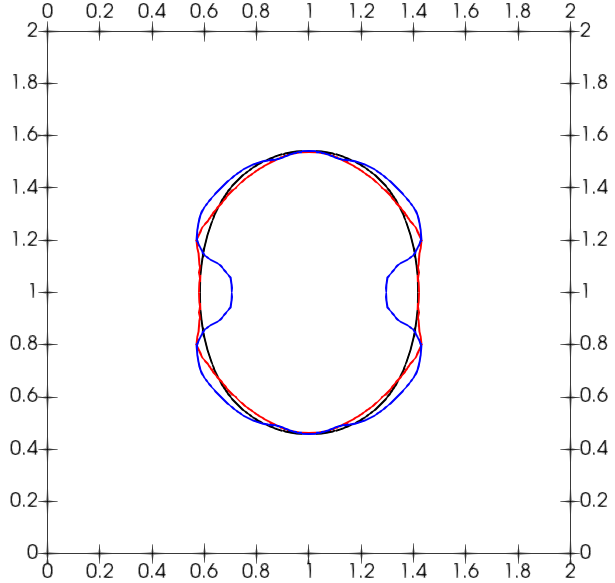}
	\caption{Oscillating bubble test case, isoline $\phi = \frac{1}{2}$ at $t = T_{f}$ with $N_{el} = 40$. The black line shows the results with the use of the Laplace-Beltrami operator \eqref{eq:surface_tension_laplace_beltrami}, the red line reports the results obtained with the use of the Bonnet's formula \eqref{eq:surface_tension_level_set}, whereas the blue line represents the results achieved using the evolution equation for the mean curvature \eqref{eq:mean_curvature_evolution}-\eqref{eq:mean_curvature_evolution_bis} to compute \eqref{eq:surface_tension_evolution_equation}. $N_{el}$ denotes the number of elements along each direction.}
	\label{fig:oscillating_bubble_phi_40x40}
\end{figure}

\begin{figure}[h!]
	\centering
	\includegraphics[width = 0.8\textwidth]{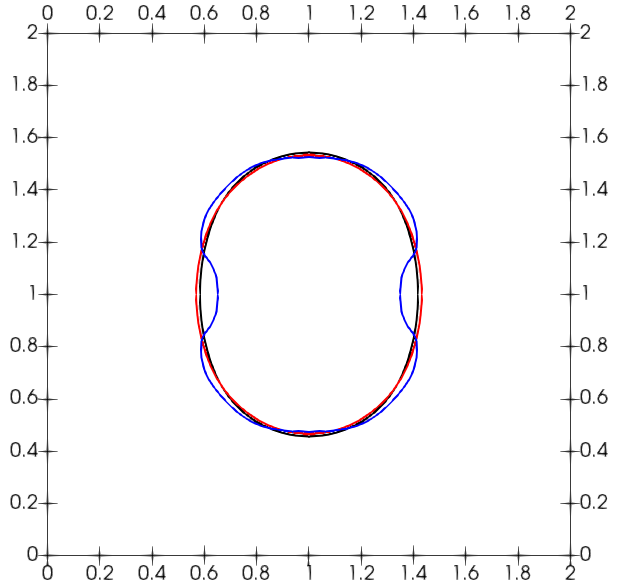}
	\caption{Oscillating bubble test case, isoline $\phi = \frac{1}{2}$ at $t = T_{f}$ with $N_{el} = 80$. The black line shows the results with the use of the Laplace-Beltrami operator \eqref{eq:surface_tension_laplace_beltrami}, the red line reports the results obtained with the use of the Bonnet's formula \eqref{eq:surface_tension_level_set}, whereas the blue line represents the results achieved using the evolution equation for the mean curvature \eqref{eq:mean_curvature_evolution}-\eqref{eq:mean_curvature_evolution_bis} to compute \eqref{eq:surface_tension_evolution_equation}. $N_{el}$ denotes the number of elements along each direction.}
	\label{fig:oscillating_bubble_phi_80x80}
\end{figure}

We now consider a more dynamic example. Starting from the configuration described in Section \ref{ssec:static_bubble}, we modify the initial shape from a circle to an ellipse by scaling the semi-axes by a factor $1.25$ in the $x$-direction and $0.75$ in the $y$-direction, respectively. The final time is now $T_{f} = \SI{20}{\second}$. Indeed, as already discussed in Section \ref{ssec:static_bubble}, for longer simulation times the approach based on the evolution equation \eqref{eq:mean_curvature_evolution}-\eqref{eq:mean_curvature_evolution_bis} is corrupted. We consider two computational grids, composed by $N_{el} = 40$ and $N_{el} = 80$ elements, along each direction, respectively. We compare the use of the Laplace-Beltrami operator \eqref{eq:surface_tension_laplace_beltrami}, the use of the Bonnet's formula \eqref{eq:surface_tension_level_set} and the computation of surface tension contribution \eqref{eq:surface_tension_evolution_equation} using the evolution equation for the mean curvature \eqref{eq:mean_curvature_evolution}-\eqref{eq:mean_curvature_evolution_bis}. Figure \ref{fig:oscillating_bubble_phi_40x40} shows the interface $\phi = \frac{1}{2}$ at the final time. One can easily notice that oscillations around the configuration captured by \eqref{eq:surface_tension_laplace_beltrami} are present using Bonnet's formula \eqref{eq:surface_tension_level_set}. These oscillations are stronger employing the evolution equation for the mean curvature \eqref{eq:mean_curvature_evolution}-\eqref{eq:mean_curvature_evolution_bis}. Analogous considerations hold using the $80 \times 80$ grid, even though the oscillations are reduced as long as the resolution increases, as one can notice from Figure \ref{fig:oscillating_bubble_phi_80x80}.

\section{Conclusions}
\label{sec:conclu}

We have performed a quantitative assessment of different strategies to compute the contribution due to surface tension in immiscible incompressible flows. The conservative level set method developed in \cite{orlando:2024} has been employed for this comparison. The results show that the use of an evolution equation for the curvature can be considered as a valid alternative to investigate in order to compute this quantity. The most accurate results presented in Section \ref{ssec:static_bubble} are those obtained using the evolution equation for the mean curvature \eqref{eq:mean_curvature_evolution}-\eqref{eq:mean_curvature_evolution_bis} to compute the surface tension contribution \eqref{eq:surface_tension_evolution_equation}. Issues are instead present for longer times and for dynamic configurations, as discussed in Section \ref{ssec:oscillating_bubble}, for which this approach, at the current stage, yields less accurate results and requires therefore further investigation. In future work, we aim to develop suitable reinitialization techniques for the evolution of the mean curvature, so as to improve the accuracy for longer simulation times and to incorporate this relation and the analogous one for the interfacial area density described in \cite{orlando:2023} into classical models for both incompressible and compressible two-phase flows, so as to improve the computation of interfacial source term in the case of a not well resolved interface.

\section*{Acknowledgments}

This work has been partially supported by the ESCAPE-2 project, European Union’s Horizon 2020 Research and Innovation Programme (Grant Agreement No. 800897).
 
\bibliographystyle{plain}
\bibliography{SurfaceTension_Curvature.bib}

\end{document}